\documentclass[11pt,twoside]{article}
\usepackage{cozumel2005}
\usepackage{epsf}
\usepackage{psfig}
\usepackage{lscape}
\def\ltsima{$\; \buildrel < \over \sim \;$}
\def\simlt{\lower.5ex\hbox{\ltsima}}
\def\gtsima{$\; \buildrel > \over \sim \;$}
\def\simgt{\lower.5ex\hbox{\gtsima}}
\begin{document}
\title{Galactic Stellar Populations}
\author{Rosemary F.G.~Wyse} 
\affil{Department of Physics \& Astronomy, Johns Hopkins University, USA}    
\author{Gerard Gilmore}
\affil{Institute of Astronomy, Cambridge University, UK}    

\begin{abstract} 

The history of the formation and evolution of the Milky Way Galaxy is
found in the spatial distribution, kinematics, age and chemical
abundance distributions of long-lived stars.  From this fossil record
one can in principle extract the star formation histories of different
components, their chemical evolution, the stellar Initial Mass
Function, the merging history -- what merged and when did it merge? --
and compare with theoretical models.  Observations are driving models,
and we live in exciting times.

\end{abstract}


\section{Context}

The large scale structure of the Universe, as probed by the Cosmic
Microwave background, provides strong evidence for a flat geometry,
$\Omega_{tot} = 1$ (e.g.~Spergel et al.~2003).  The motions of
galaxies constrain the matter contribution to be $\Omega_{matter} \sim
0.3$, while Big Bang Nucleosynthesis calculations provide an upper
limit to the baryon content of the Universe of $\Omega_{baryon} \simlt
0.05$ (e.g.~Schramm 2000). These discrepant values are understood with
the postulate of non-baryonic (dark) matter, and the realization that
the usual decelerated expansion of matter-dominated universes does not
describe the recent (redshifts $\simlt 0.5$) expansion of our
Universe, which is instead {\it accelerating\/} its rate of expansion.
This acceleration indicates the recent dominance of a component with
negative pressure, most commonly and simply ascribed to a Cosmological
Constant/Vacuum Energy/`Lambda-term', which has constant energy
density as the Universe expands and at the present epoch provides
$\Omega_\Lambda \sim 0.7$.  The dark matter is most probably `cold'
i.e.~was nonrelativistic when it decoupled from the other constituents
of the Universe, resulting in the preservation of primordial
fluctuations on all scales from sub-Galactic up.

Galaxy formation and evolution in such a `$\Lambda$CDM' Universe
starts with the collapse of perhaps $\sim 10^6{\cal M}_\odot$
overdense regions (the characteristic mass of `the first objects' is
the subject of much on-going work) and large galaxies like the Milky
Way emerge as the end-point of a process of hierarchical clustering,
merging and accretion. Abadi et al.~(2003) present a recent simulation
of the formation of a present-day disk galaxy that demonstrates many
of the important aspects, including the outstanding problem of how to
include star formation and gas physics.  Generic predictions for disk
galaxies include the following:

\begin{itemize}

\item{} Extended disks form late, after a redshift of unity, or a
lookback time of $\sim 8$~Gyr, in order to avoid losing too much
angular momentum during active merging at earlier times

\item{} A large disk galaxy should have hundreds of surviving
satellite dark haloes at the present day

\item{} The stellar halo is formed from disrupted satellites  

\item{} Minor mergers (a mass ratio of $\sim 10- 20$\% between the
satelite and the disk) into a disk heat it, forming a thick disk out
of a pre-existing thin disk, and create torques that drive gas into
the central (bulge?) regions

\item{} More significant mergers transform a disk galaxy into an S0 or
even an elliptical

\item{} Subsequent accretion of gas can reform a thin disk 

\item{} Stars can be accreted into the thin disk from suitably massive
satellites (dynamical friction must be efficient) and if to masquerade
as stars formed in the thin disk, must be on suitable high angular
momentum, prograde orbits

\end{itemize}

The merging history of a typical massive-galaxy dark halo is fairly
straightforward to calculate, since only gravity is involved. However,
most simulations lack the resolution to follow how far inside a
`parent' halo a merging satellite penetrates, and this is crucial to
determine the effect on the baryonic disk. Semi-analytic techniques
have been developed (e.g.~T\'oth \& Ostriker 1992; Taylor \& Babul
2004; Benson et al.~2004; Zentner et al.~2005) but these involve
further uncertainties, the treatment of dynamical friction and tidal
effects being crucial issues. Of course, inclusion of the baryons, as
stars and gas, is more uncertain.  Dealing solely with dissipationless
physics, i.e.~pure stellar baryonic components plus dark matter,
N-body simulations have shown that a merger between a disk galaxy and
a robust (fairly dense) satellite of even 10\% of the mass of the
disk, and thus even a smaller fraction of the whole galaxy, $< 2\%$ in
the models, is sufficient to result in significant thickening, as
shown by Velaquez \& White (1999) and by Walker, Mihos \& Hernquist
(1996).  High-resolution cosmological simulations predict that a
galaxy of the mass of the Milky Way now would typically have
experienced of order 3 mergers with satellite systems of mass at least
10\% of that of the total galaxy in the last 10~Gyr (A.~Berlind,
private communication). Many more mergers with smaller mass ratios,
which from the simulations would be capable of thickening a thin disk,
would be expected.  As we discuss below, the thick disk of the Milky
Way is apparently composed of unifomly old stars, ages $\simgt
10$~Gyr, implying that the thin disk of the Milky Way has not been
heated by mergers over the last 10~Gyr.  Consistency between these
remains to be found, unless we are happy with a special place in the
Universe.

$\Lambda$CDM models predict about a factor of ten more disrupted
satellite galaxies than surviving (see Fig.~1 of Bullock, Kravtsov \&
Weinberg 2000), providing a total stellar mass to the field halo of
$\simgt 5 \times 10^8{\cal M}_\odot$, comparable to the estimated
stellar mass of the entire halo ($\sim 10^9{\cal M}_\odot$, Carney et
al.~1990).  Most disruption occurs `relatively early, $z \simgt 0.5$'
(Bullock et al.~2000), or a lookback time of $\simgt 5.5$~Gyr.  One
might still expect to see signatures of these hundred or so disrupted
dwarfs, since simulations suggest that signatures in phase space,
particularly if integrals of the motion can be estimated, can survive
for $\sim$ a Hubble time.

\section{The Milky Way Galaxy: Ongoing Mergers}

The Milky Way is clearly merging with at least one satellite galaxy at
the present day, namely the Sagittarius dwarf spheroidal (Ibata,
Gilmore \& Irwin 1994). This tidal interaction has produced, and is
producing, thin streams, or tidal arms, that can be traced across the
sky (e.g.~Majewski et al.~2003) -- and that can in principle be used
to constrain the shape and lumpiness of the Galactic dark halo.  The
mass ratio of this merger is far from clear -- the present stellar
mass is $\sim 3 \times 10^7{\cal M}_\odot$ and present total mass is
inferred from internal kinematics and models of the tidal debris to be
$\sim 5 \times 10^8{\cal M}_\odot$ (e.g.~Law, Johnston \& Majewski
2005 and references therein).  The fact that the Sagittarius dwarf
apparently self-enriched to approximately solar metallicity at least a
Gyr ago (Bonifacio et al.~2000; McWilliam \& Smecker-Hane 2005; Monaco
et al.~these proceedings) suggests a potential well deeper than that
of the LMC, which has yet to reach this level of enrichment. Models of
tidal disruption, over a 13~Gyr time period, in a fixed potential
similar to that of the present Milky Way suggest a mass in tidal
debris now that is some five times larger than that remaining bound,
i.e.~an initial mass around 6 times the present one (e.g.~Helmi \&
White 2001) but this is not a very realistic model in the context of
$\Lambda$CDM, and the results are sensitive to the assumed density
profile of the satellite.  It is also not clear when the Sgr dSph
achieved its present orbit, and how that would be related to `merger
epoch' as defined in theoretical models.  More work is clearly
needed. In any case, interactions with the Sgr dSph may well be
playing a role in driving/maintaining the complex structure of the
outer disk (e.g.~Ibata \& Razoumov 1998).

Tidal arms, combining spatial and kinematic substruture, provide the
cleanest evidence of interactions.  However, the beautiful arms of the
outer halo globular cluster Pal 5 (Odenkirchen et al.~2003),
identified through the excellent photometry from the Sloan Digital Sky
survey and extending over 10 degrees on the sky, serve to remind us
that streams are not necessarily a sign of accretion and mergings.

The ambiguity between interactions and mergers may be illustrated by
the LMC/SMC/Magellanic stream system. The relative roles of tidal
effects and gaseous, ram-pressure stripping in this system remain
unclear, with even the leading gaseous arm (Putman et al.~1998)
perhaps being reproduced by gas physics (Mastropietro et al.~2005),
albeit with associated shrinking of the periGalacticon of the LMC by
$\sim 10\%$ per ($\sim 2$Gyr) orbit and significant tidal stripping of
dark matter from the LMC halo.

\section{The Stellar Populations and Evolution of the Thin Disk}

As noted above, extended thin disks are constrained in $\Lambda$CDM to
form after the bulk of merging is complete, $z \simlt 1$ or lookback
times of $\simlt 8$~Gyr, in order to avoid losing too much angular
momentum (e.g.~Navarro, Frenk \& White 1995; Navarro \& Steinmetz
1997; Eke, Efstathiou \& Wright 2000).

Setting aside for the moment the thick disk and the constraints that
the old ages of its constituent stars provide on the epoch of
formation of its possible thin disk progenitor, one can estimate the
epoch of the onset of star formation in the local thin disk by various
means such as from isochrone-fitting of the Hipparcos colour-absolute
magnitude diagram; isochrone fitting to Stromgren photometry of nearby
stars; the distribution of atmospheric activity age indicators, and
from the local white dwarf luminosity function.  Each technique has
its own limitations and uncertainties, but it appears that the local
disk has no shortage of stars older than 8~Gyr (e.g.~Nordstrom et
al.~2004; Binney, Dehnen \& Bertelli 2000).

The scale-length of old stars in the thin disk is 2--4~kpc
(e.g.~Siegel et al.~2002), so that the solar neighbourhood is some
$\sim 3$~scalelengths from the center. Thus, provided the old stars
locally were born locally, the formation of an extended disk in the
Milky Way was {\it not\/} delayed until after a redshift of unity.
This is a potential problem for $\Lambda$CDM models, particularly
since M31 apparently also has an extended disk of old stars (Ferguson
\& Johnston 2001). Furthermore, direct observation of high redshift
galaxies, out to a redshift $z \sim 3$, in the rest-frame optical
(Trujillo et al.~2005), reveals little evolution in disk size,
significantly less than predicted by semi-analytic CDM models (Mo, Mao
\& White 1999), but consistent with the simplest picture of gaseous
infall and star formation within a fixed potential well, with the star
formation rate higher in the central disk.  Indeed, the interpretation
from these high redshift observations is that `stellar disks form from
early on, in large haloes' (Trujillo et al.~2005).

An alternative interpretation of the old stars in the local disk is to
posit that they formed somewhere else and were deposited in the local
disk at some subsequent time.  There has been much recent activity in
disk dynamics and mixing (e.g.~Fuchs 2001; Sellwood \& Binney 2002; De
Simone, Wu \& Tremaine 2004), an aspect of which is how much net
migration of stars outward across the disk is possible; at present the
results are not clear.  The possibility that a significant fraction of
the old thin disk could have been accreted as debris from a few
satellite galaxies was raised by Abadi et al.~(2003).  A typical
satellite orbit is far from circular (e.g.~Benson 2005) so that this
scenario requires that the satellites be massive enough for dynamical
friction to damp vertical motions and circularize their orbit quickly
enough, and it remains to be seen if, for example, the chemical
composition of the old disk stars is consistent.

Ongoing large-scale spectroscopic surveys such as RAVE (targeting
bright stars with the UKSchmidt Telescope and 6dF multi-object
spectrograph; Steinmetz 2003) and the Galactic structure survey of the
Sloan Digital Sky Survey Extension (SDSS-II/SEGUE) should provide
ideal databases for an identification of kinematic substructure in the
disk.  The high-resolution mode of the proposed multi-object
spectrograph (KAOS/WFMOS) for Gemini will provide unprecedented
elemental abundance data, containing much more information than
overall metallicity, and with signatures that persist longer than
spatial or even most kinematic features.

The `ring' around the Galaxy seen in star counts (e.g.~Yanny et
al.~2003; Ibata et al.~2003) could be either a remnant of satellite
accretion into the plane of the disk (e.g.~Bellazzini et al.~2004;
Martin et al.~2004; Rocha-Pinto et al.~2005), or more simply structure
in the outer stellar disk, which most probably warps and flares
(e.g.~Momany et al.~2004).  Indeed the rich structure in HI gas in the
outer disk may be seen in the Leiden/Argentine/Bonn HI survey
(Kalberla et al.~2005). Even the old disk is unlikely to be well-fit
by a smooth model, given the strong spiral structure seen in K-band
images of external spirals (Rix \& Zaritsky 1995).  Again, ongoing
surveys will provide much-needed information.

However, the interpretation of large datasets of stellar kinematics
and metallicity in the context of searches for substructure in the
disk is complicated by the fact that the underlying potential of the
disk is neither smooth, axisymmetric nor time-independent.  As
demonstrated by De Simone, Wu \& Tremaine (2004), transient
perturbations, such as segments of spiral arms, not only heat the
stellar disk, but produce `moving groups' that persist long after the
gravitational perturbation has gone.  These kinematic features are
created from random collections of disk stars, and so will contain a
range of ages and metallicities.  An unwary astronomer might conclude
that such a complex stellar population indicates disruption of a
satellite galaxy.

Analysis of the K/M giants with accurate radial velocities and
Hipparcos distances and proper motions has indeed identified several
kinematic moving groups with members of a range of ages and/or
metallicities (Famaey et al.~2004), which those authors have ascribed
to dynamical perturbations within the disk.  Helmi et al.~(2005)
analysed the rich dataset from Nordstrom et al.~(2004), consisting of
full 3-D space motions (from Hipparcos parallaxes, Tycho-2 proper
motions, and new multi-epoch radial velocities) and effective
temperatures and metallicities derived from {\it uvby}$\beta$ Str\"
omgren photometry.  They concluded that while some substructure was
due to the bar potential and transient spiral arms, there remained
substructures that they identified as having metallicity and age
distributions more consistent with debris from accreted and disrupted
satellites.  It will be very interesting to cross-correlate these
latter substructures with the spectroscopic metallity and elemental
abundance dataset of Soubiran \& Girard (2005), to investigate their
detailed abundance patterns.

\section{The Stellar Populations and Evolution of the Thick Disk}

The thick disk was defined through star counts 20 years ago (Gilmore
\& Reid 1983) and is now well-established as a distinct component, not
the tail of the stellar halo or of the thin disk.  Its origins remain
the source of considerable debate.  Locally, some $\sim 5\%$ of stars
are in the thick disk; the vertical scale-height is $\sim 1$~kpc, and
radial scale-length $\sim 3$~kpc.  Assuming a smooth
double-exponential spatial distribution with these parameter values,
the stellar mass of the thick disk is 10--20\% of that of the thin
disk (the uncertainty allowing for the uncertainty in the structural
parameters), or some $10^{10} \cal{M}_\odot$.  Broadly similar structures
have been seen in the resolved stellar populations of nearby spirals
(e.g.~Mould 2005; Yoachim \& Dalcanton 2005).

Again the properties of the stellar populations in this Galactic
component are rather poorly known far from the solar
neighborhood. Locally, within a few kpc of the Sun, the typical thick
disk star is of intermediate metallicity, ${\rm [Fe/H] \sim
-0.6}$~dex, and old, with an age comparable to that of 47~Tuc, the
globular cluster of the same metallicity, $\sim 12$~Gyr (see
e.g.~review of Wyse 2000).  Detailed elemental abundances are now
available for statistically significant sample sizes. These show that
the pattern of elemental abundances differs between the thick and thin
disks, with different values of the ratio [$\alpha$/Fe] at fixed
[Fe/H], implying distinct star formation and enrichment histories for
the thick and thin disks (Fuhrmann 1998, 2004; Prochaska et al.~2000;
Feltzing, Bensby \& Lundstr\"om 2003; Nissen 2004; Bensby et
al.~2005).  The data show a `Type~II supernova plateau' for metal-poor
thick disk stars, ${\rm [\alpha/Fe]} \sim +0.4$ for ${\rm [Fe/H]}
\simlt -0.5$, with a downturn for more metal-rich stars (see Figure~1,
taken from Bensby, Feltzing \& Lundstr\"{o}m 2004), indicating
contributions to iron from the longer-lived Type Ia supernovae.  The
fact that the Type~II plateau has the same value as that seen in stars
belonging to the stellar halo implies that these two populations were
enriched by Type~II supernovae from massive stars of the same Initial
Mass Function (see e.g.~Gilmore \& Wyse 1998 for more discussion of
this point).  The turn-down implies that star formation in the
progenitor of the thick disk persisted long enough for Type Ia
supernovae to explode and for their iron to be incorporated in its
chemical evolution; models imply this is a duration of 1--2~Gyr
(e.g.~Matteucci \& Greggio 1986; Smecker-Hane \& Wyse 1992).

\vskip -6 truecm
\begin{figure}[ht!]
\begin{center}
\psfig{figure=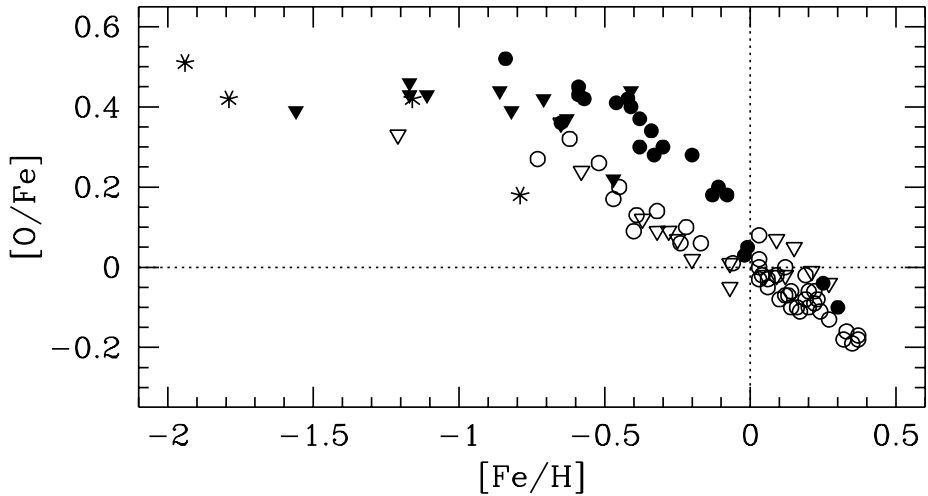,width=5.in}
\caption{Taken from Bensby, Feltzing \& Lundstr\"{o}m 2004, their
Figure~11.  Filled symbols represent stars whose kinematics are
consistent with membership of the thick disk, while open symbols
represent thin disk stars.  At a given value of [Fe/H], the thick and
thin disk stars are separated, with thick disk stars having higher
[O/Fe].  At the typical thick disk metallicity, [Fe/H]$ \sim
-0.5$~dex, the value of [O/Fe] in thick disk stars is equal to that
seen in the stellar halo, and consistent with enrichment by Type II
supernovae.  More metal-rich thick disk stars show some enrichment by
iron-dominated ejecta from Type Ia supernovae.}
\end{center}
\end{figure}

The kinematics of the thick disk are intermediate between those of the
thin disk and those of the stellar halo; in particular, the standard
value for the mean azimuthal streaming velocity of the thick disk is
$V_{rotation} \sim 170$~km/s (Norris 1986; Morrison, Flynn \& Freeman
1990; Chiba \& Beers 2000). However, surveys of faint F/G stars tend
to find a lower value, $V_{rotation} \sim 100$~km/s (e.g.~Wyse \&
Gilmore 1986; Gilmore, Wyse \& Norris 2002) and we return to the
interpretation of this below.

The fairly high values for the velocity dispersions of the thick disk,
$\sigma_W \sim 40$~km/s and $\sigma_{total} \sim 80$~km/s, argue
against normal disk-heating mechanisms (the transient gravitational
perturbations in the disk, bits of spiral arms, GMCs..) being involved
in its formation -- those processes generally saturate at the values
of the velocity dispersions for the old thin disk, or $\sigma_W \sim
20$km/s.  Exceptional heating of the thin disk that occurs over cosmic
time, such as would happen if the dark matter in the halo were massive
black holes (Lacey \& Ostriker 1985), is ruled out as the formation
mechanism for the thick disk by the extended star formation in the
thin disk, and the apparent lack of intermediate-age or young stars in
the thick disk.  Exceptional heating that occurs only very early in
the evolution of the thin disk is a viable possibility, and this is
what occurs in a minor merger at high redshift.  The old age of the
stars in the thick disk, $\sim 10-12$~Gyr (Gilmore, Wyse \& Jones
1995; Nordstrom et al.~2004; Fuhrmann 2004), would constrain any
merger violent enough to form a thick disk from the thin disk to have
happened at this look-back time, or redshift $z \simgt 1.5$. Given
that, as noted above, models have found that a merger with as little
as 10\% of the {\it disk} mass, or a few percent of the dark halo
mass, is sufficient to form a thick disk (e.g.~Velaquez \& White
1999), and that cosmological calculations imply that such low mass-ratio
mergers of dark haloes continue to later times, it is clearly important
to (i) obtain reliable, precise and accurate ages for large samples of
thick disk stars (large samples are needed to isolate contaminants,
such as thin disk stars ejected into the thick disk by binary
supernova explosions), and (ii) develop robust techniques for
modelling baryonic galaxies in the cosmological simulations.

What signatures of a merger-origin for the thick disk might remain
observable today?  In the merger, orbital energy is deposited in the
internal degrees of freedom of both the thin disk and the satellite,
and acts to disrupt the satellite and heat the disk.  Depending on the
orbit of the satallite, and on its density profile and mass (this last
determines the dynamical friction timescale), tidal debris from the
satellite will be distributed through the larger galaxy during the
merger process.  Thus the phase space structure of the debris from the
satellite depends on many parameters, but in general one expects that
the final `thick disk' will be a mix of heated thin disk and satellite
debris.  The age and metallicity distributions of the thick disk can
provide constraints on the mix.

Could the thick disk be dominated by the debris of tidally disrupted
dwarf galaxies (cf.~Abadi et al.~2003)?  As noted above, the local
(within a few kpc of the Sun) thick disk is old and quite metal-rich,
with a mean iron abundance $\sim -0.6$~dex.  Further, the bulk of
these stars have enhanced, super-solar [$\alpha$/Fe] abundances (see
Figure~1).  Achieving such a high level of enrichment so long ago (the
stellar age equals the age of 47~Tuc, at least 10~Gyr, as noted
above), in a relatively short time -- so that Type II supernovae
dominate the enrichment, as evidenced by the enhanced levels of
[$\alpha$/Fe]) -- implies a high star formation rate within a rather
deep overall potential well.  This does not favor dwarf galaxies.

Indeed, the inner disk of the LMC, our present most massive satellite
galaxy, has a derived metallicity distribution (Cole, Smecker-Hane \&
Gallagher 2000) that is similar to that of the (local) thick disk,
but, based on the color-magnitude diagram, these stars are of
intermediate age (see also Hill et al.~2000). Thus the LMC apparently
took until a few Gyr ago to self-enrich to an overall metallicity that
equals that of the typical local thick disk star in the Galaxy, which
is much older.  Further, the abundances of the $\alpha$-elements to
iron in such metal-rich LMC stars are below the solar ratio (Smith et
al.~2002), unlike the local thick disk stars.  This may be understood
in terms of the different star-formation histories (cf.~Gilmore \&
Wyse 1991).  The LMC is not a good template for a putative dwarf that
could have been tidally disrupted to form the thick disk from its
debris.
 
None of the retinue of dSph satellite galaxies has a stellar
population well-matched to that of the thick disk, which requires old
age, enhanced values of ${\rm [\alpha/Fe]}$ for metallicities below
$\sim -0.5$~dex, and overall higher mean metallicity than a typical
dSph (e.g.~Tolstoy et al.~2003).  One is left with the conclusion that
if a disrupted dwarf forms the bulk of the thick disk, then that
satellite was very different from those that survived (we will echo
this conclusion below, in the discussion of the stellar halo).
 
However, it is very possible -- and indeed, required in merger-models
-- that a significant minority of `thick disk' stars are debris from
accreted satellite(s). These stars would be expected to be more
metal-poor than the bulk of the (true) thick disk stars, and to be on
orbits more similar to that of a typical satellite, than the
high-angular momentum orbit of a typical thick disk star. In
$\Lambda$CDM cosmological models the typical satellite/subhalo initial
orbital angular momentum is close to half that of a circular orbit of
the same energy (e.g.~Benson 2005; Zentner et al.~2005) so one would
expect an azimuthal streaming velocity of around 100km/s for satellite
debris in a system like the Milky Way with a flat rotation (circular
velocity) curve, with amplitude $\sim 220$~km/s.  Of course in
hierarchical clustering models the potential well of the Milky Way is
not fixed, but theory predicts an early epoch of massive mergers that
sets the potential well depth (e.g.~Zentner \& Bullock 2003).

A component with a mean azimuthal streaming velocity of around
100~km/s is indeed seen in the radial velocity datasets of faint ($V
\simgt 18$) F/G dwarfs of Gilmore, Wyse \& Norris (2002) and Wyse et
al.~(2005), in widely separated lines-of-sight, and in datasets taken
with a variety of multi-object spectrographs and telescopes.
Reassuringly, hints are also seen in the K-giants of Morrison, Flynn
\& Freeman (1990, their Figure~7(d) and (g)) and in the local sample
of Beers et al.~(2000, as shown by Navarro, Helmi \& Freeman~2004).

We interpret this as strong evidence in favour of the merger-heated
disk origin for the thick disk.

\section{The Stellar Populations and Evolution of the Stellar Halo}

\subsection{The Field Stellar Halo}

The total stellar mass of the halo is $\sim 2 \times 10^9
\cal{M}_\odot$ (cf.~Carney, Laird \& Latham 1990), modulo
uncertainties in the stellar halo density profile in each of the outer
halo, where substructure may dominate, and the central regions, where
the bulge dominates.  Some $\sim 30\%$ of the stars in the halo are on
orbits that take them through the solar neighborhood, to be identified
by their `high-velocity' with respect to the Sun.  These stars form a
rather uniform population -- old and metal-poor, with enhanced values
of the elemental abundance ratio [$\alpha$/Fe] (see e.g.~Unavane, Wyse
\& Gilmore 1996).  The dominant signature of enrichment by Type~II
supernovae indicates a short duration of star formation, so that the
ejecta of the longer-lived Type~Ia supernovae is not incorporated into
stars.  This could naturally arise due to star formation and
self-enrichment occuring in low-mass star-forming regions that cannot
sustain extended star formation.  Indeed, the stellar density of the
halo is sufficiently low that its stars must have formed in higher
density systems and then been dispersed, given the observational and
theoretical constraints on the location and environments of star
formation -- i.e.~that star formation is limited to cold, dense gas
clouds.  Many theoretical and observational studies have suggested
that star formation occurs in clusters and associations of a range of
masses and binding energies, and a large fraction of the stellar
structures formed are either never bound (after gas dispersal) or are
disrupted on several crossing times by a combination of internal and
external forces (e.g.~review of Fall 2004). The field stellar halo
could rather naturally be formed from the debris from such stellar
`clusters', with the on-going mass-loss from e.g.~Pal 5 (Odenkirchen
et al.~2003) the present-day manifestation of a continual and
continuing process.

The mean metallicity of the stellar halo, around $-1.5$~dex (e.g.~Ryan
\& Norris 1991), is low compared with that of the solar neighborhood,
which is around $-0.2$~dex (Nordstrom et al.~2004).  With a fixed
stellar initial mass function, and no gas flows, one expects a system
of higher gas fraction, such as the local disk, to be {\it less\/}
chemically evolved than a system with lower gas fraction, such as the
stellar halo.  Hartwick (1976) provided an elegant explanation that
related low mean metallicity to gas outflows during star formation;
this is expected in scenarios of the formation of the stellar halo,
such as just described, that postulate star-formation in low binding
energy systems.  The chemical evolution requirements are such that for
a fixed stellar IMF, the outflows must occur at around 10 times the
rate of star formation.  An attractive corollary to this picture is
that one can tie the gas outflow from halo star-forming regions to gas
inflow to the central regions to form the bulge; the low angular
momentum of halo material means that it will only come into
centrifugal equilibrium after collapsing in radius by a significant
factor.  The mass ratio of bulge to halo is around a factor of ten,
just as would be expected, and the specific angular momentum
distributions of stellar halo and bulge match (Wyse \& Gilmore 1992;
see Figure~2 here).

\begin{figure}[h!]
\begin{center}
\psfig{figure=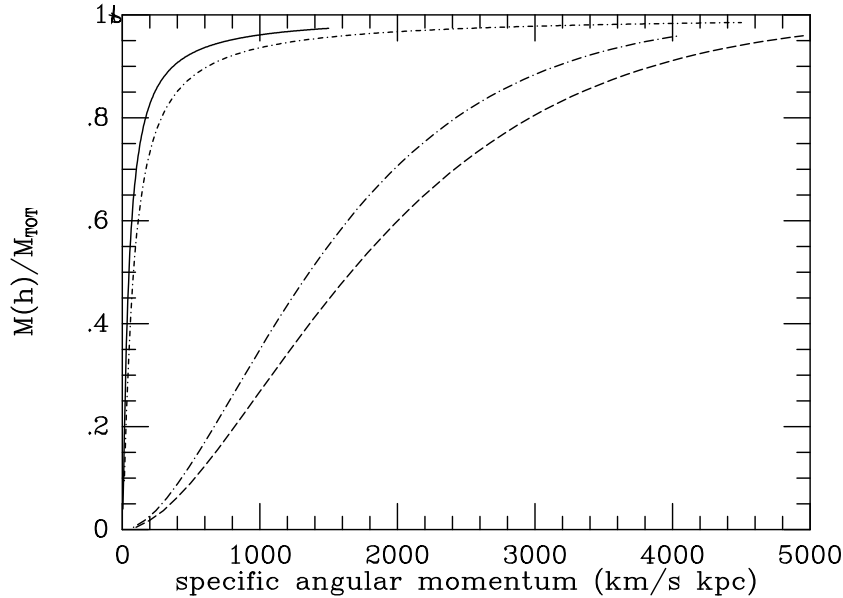,width=4.5in}
\caption{Adapted from Wyse \& Gilmore 1992, their Figure~1.  Angular
momentum distributions of the bulge (solid curve), the stellar halo
(short-dashed/dotted curve), the thick disk (long-dashed/dotted curve)
and the thin disk (long-dashed curve).  The bulge and stellar halo
have similar distributions, as do the thick and thin disks.}
\end{center}
\end{figure}

Indeed, the field stellar halo can be formed through the disruption of
any flavour of system, as long as it formed stars only very early in
cosmic time (to be compatible with the old age of the halo) and did
not self-enrich significantly (to be consistent with the low mean
overall metallicity plus the apparent lack of enrichment by Type ~Ia
supernovae).  Hierarchical clustering cosmologies such as $\Lambda$CDM
postulate that the stellar halo formed through the disruption of small
galaxies; the only example of a viable satellite galaxy for this, at
the present day, is the Ursa Minor dwarf spheroidal (dSph), being the
only one with a uniformly old, metal-poor population (e.g.~Unavane et
al.~1996). And even for this dSph, the pattern of elemental ratios may
not match those in the stellar halo (e.g.~Shetrone et al.~2003;
Tolstoy et al.~2003).  A typical star in a typical dSph, which has had
star formation over an extended period, longer than in the Ursa Minor
dSph, lacks the enhanced level of [$\alpha$/Fe] seen in the field halo
stars (Venn et al.~2004).  Indeed, analysis of the age distributions
of stars in dSph, and in the stellar halo, as a function of
metallicity (Unavane et al.~1996) demonstrates that systems like the
extant dSph could not form the bulk of the stellar halo unless they
were accreted more than $\simgt 8$~Gyr ago and further star formation
was rapidly truncated after accretion.  A similar conclusion, that
extent dSph could not provide a significant fraction of stars in the
stellar halo, follows from analysis of their elemental abundances
(Venn et al.~2004).  The response of $\Lambda$CDM advocates is to
argue that the `satellites' that formed the stellar halo were indeed
accreted only very early, and that this early accretion necessarily
makes those satellites different from the survivors, traced by the
dSph, which were accreted later (e.g.~Bullock \& Johnston 2005). The
required truncation of star formation for early accretions, but not
for later accretion (e.g.~LMC analogs), is simply assumed
(e.g.~Robertson et al.~2005).

A further constraint is the remarkably low level of cosmic (i.e.~not
observational uncertainty) scatter seen in the ratios of
$\alpha$-elements to iron in halo stars (e.g.~Cayrel et al.~2004), as
low as $\sim 0.05$~dex. This requires a fixed invariant massive-star
IMF (e.g.~Wyse \& Gilmore 1992; Wyse 1998) and little `cross-talk'
between star-forming regions (Gilmore \& Wyse 1998; see also Fran\c
cois et al.~2004).
  
Of course, as noted above, mergers and accretion from satellite
galaxies do play some role in the evolution of the Milky Way.  As
kinematic and metallicity datasets increase in size and quality
structure is being identified, such as a retrograde stream (Gilmore,
Wyse \& Norris 2002) that may be associated with the globular cluster
$\omega$~Cen, or at least the posited parent dwarf galaxy of which
this globular cluster would be the surviving nucleus (Mizutani, Chiba
\& Sakamoto 2003).  Streams are more rare in the inner halo ($R_G
\simlt R_{solar\, circle}$, which contains most of the stellar mass),
perhaps reflecting shorter dynamical timescales.  Moving groups in
local high-velocity stars have indeed been isolated (e.g.~Helmi et
al.~1999; Meza et al.~2005), but mass estimates are uncertain when
based on few stars (see Chiba \& Beers 2000), as are their origins --
perhaps even local streams are associated with the Sagittarius dwarf
(e.g.~Majewski et al.~2003), or the putative parent of $\omega$~Cen
(Meza et al.~2005).

\begin{figure}[h!]
\begin{center}
\psfig{figure=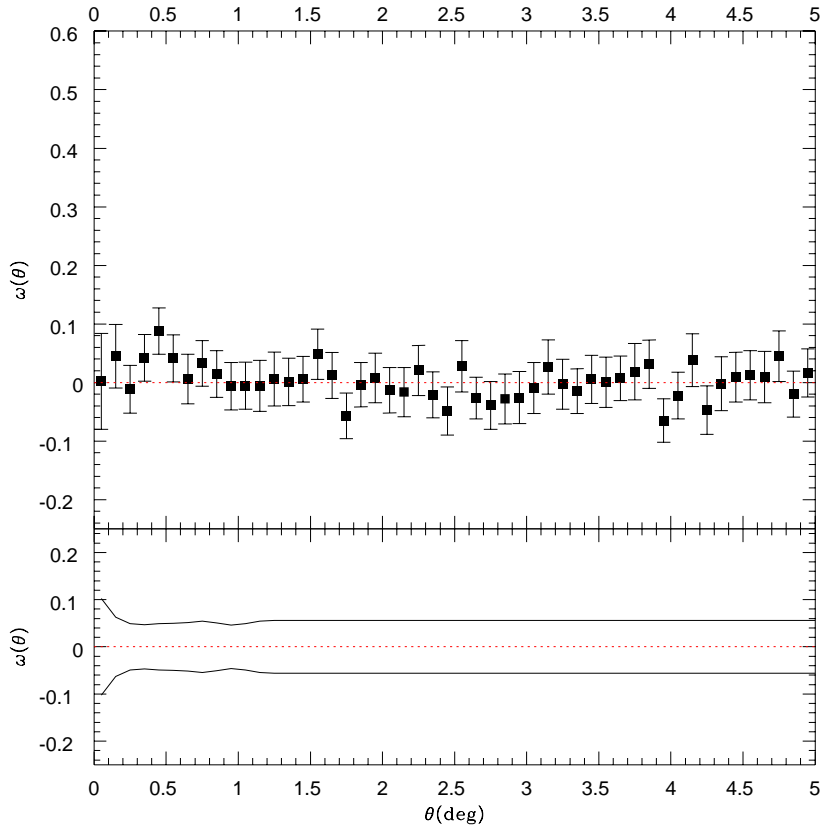,width=4.5in}
\caption{Taken from Lemon, Wyse, Liske, Driver \& Horne 2004, their
Figure 12.  The upper panel shows the 2-point correlation function for
F-stars brighter than B=19 in the $\sim 37.5$~square degrees imaged
for the Millenium Galaxy Catalogue.  The mean correlation function is
flat and essentially zero; the errors shown are $\pm 1\sigma$. The
lower panel shows the result of 200 simulations of random samples,
taking into account the overall stellar gradient across the sample due
to the large-scale structure of the Galaxy, with the solid lines
giving the $1\sigma$ range.  The data are consistent with no
small-scale correlations.}
\end{center}
\end{figure}

More recent accretion into the stellar halo may be constrained by
spatial structure, or lack thereof.  The 2-point correlation function
of relatively bright, $B < 19$, main sequence stars shows little
structure, illustrated in Figure~3 (Lemon et al.~2004), as is the case
for halo Blue Horizontal Branch stars (Brown et al.~2004).  Recent
accretion is not a dominant mechanism for formation of the bulk of the
stellar halo.  The outer halo does show structure in coordinate space,
for example in analyses of faint imaging data from the Sloan Digital
Sky Survey (Newberg et al.~2002), a significant part of which is due
to the Sgr dSph, but not all.  Kinematics and metallicity data will be
crucial in the interpretation.

Analysis of the dynamics, evolution and survival of substructure in
hierarchical models is beyond the resolution limits of most
cosmological simulations, but can be addressed by galaxy-scale
simulations and by semi-analytic means.  Interactions between the
subsystems can dominate (e.g.~Zhang et al.~2002; Penarrubia \& Benson
2004; Knebe et al.~2005) requiring rather detailed information to make
predictions.

Looking beyond the Milky Way, models for the field halo stellar
population must address the apparent correlation between the mean
metallicity of the halo stars and the host luminosity (Mouhcine et
al.~2005; see also Brodie \& Huchra 1991 for a version using globular
clusters to trace the stellar halo, updated to include only the blue,
metal-poor cluster subsystem in Strader, Brodie \& Forbes 2004).  Such
correlations, including the overall luminosity--metallicity
relationship (e.g.~Dekel \& Woo 2003) are not easily reproduceable in
hierarchical clustering models, where merger history plays a strong
role (see e.g.~Renda et al.~2005).  Correlations of the properties of
the stellar populations with overall potential well depth suggest that
if mergers are indeed the dominant mechanism by which galaxies form
and evolve, the mergers must be gas rich, and most stars form during
the mergers, with the last significant merger dominating the star
formation history.

\subsection{Globular Clusters}

The seminal paper of Searle \& Zinn (1978) proposed that the globular
clusters of the outer halo formed in `transient protogalactic
fragments' that were accreted after the formation of the central
regions.  This conclusion was based on an analysis of the available
metallicity and horizontal branch morphology data, which implied that
(assuming that age is the `second parameter') the outer clusters had a
range of ages, while the inner clusters were uniformly old. This
general age spread -- Galactocentric distance trend has been confirmed
by analyses based on deep colour-magnitude data, including the main
sequence turn-off (e.g.~Salaris \& Weiss 2002).  Furthermore, deep HST
imaging has shown similarities between the outer globular clusters and
those of satellite galaxies, in terms of both HB morphology as a
function of metallicity and values of their core radii (Mackey \&
Gilmore 2004).  Whether this indicates that the outer clusters were
acquired by the Milky Way during the accretion of systems like the
present-day satellites (Mackey \& Gilmore 2004), or that star clusters
that form and evolve in low density regions have large core radii, is
unclear (see also Wilkinson et al.~2003).  Elemental abundances of
halo stars across the Galaxy will be of interest, to investigate
whether the outer field halo has a similar pattern to the stars in
dSph, and is part of the science case of the proposed multi-object
high spectral-resolution next-generation Gemini instrument WFMOS.

The dynamical evolution of (globular) clusters is determined by a
complex mix of internal and external effects. The beautiful tidal
arms/tails from Pal 5 (Odenkirchen et al.~2003) confirm the
expectation of mass loss. It will be very interesting to re-visit,
with deep multi-band CCD photometry, the 20 globular clusters for
which Leon, Meylan \& Combes (2000) detected low surface density tidal
tails from wavelet analysis of star counts from photographic
plates. In particular, these authors found significant recent mass
loss from $\omega$~Cen, which as we noted above has been proposed as a
surviving core of an accreted dwarf galaxy (e.g.~Bekki \& Freeman
2003).

The formation and survival of (diffuse) clusters in relatively shallow
potential gradients is of particular interest in the context of dwarf
spheroidal galaxies and their retinue of globulars (setting aside the
separate puzzle of why the orbits have not decayed through dynamical
friction -- Hernandez \& Gilmore 1998; Lotz et al.~2001). Similarly to
the situation with the field stellar halo in the Milky Way, the field
population in dSph must have formed at significantly higher densities,
presumably in star clusters that were subsequently dispersed. Again,
similarly to the Milky Way halo, remnant structure may have been
identified, in the form of cold streams, in the dSph (Kleyna,
Wilkinson, Gilmore \& Evans 2003; Kleyna, Wilkinson, Evans \& Gilmore
2004).

\subsection{The Dwarf Spheroidal Galaxies}

The dSph satellite galaxies of the Milky Way have been the focus of
much recent effort, both observational and theoretical, addressing the
many ways in which these systems could interact with, and contribute
to, the Milky Way, and could impact our understanding of galaxy
formation and evolution and the nature of dark matter.  The kinematic
and photometric data that allowed the identification of the
substructure in dSph mentioned in the last subsection is part of such
a large survey, using several 8m-class ground-based
telescopes. Chemical abundance data, including individual elements,
are now available for most of the dSph (e.g.~Venn et al.~2004; Koch et
al.~2005).  The radial velocities over the face of the galaxies,
extending beyond the nominal `tidal radius', provide constraints on
the inferred dark matter distribution (e.g.~Kleyna et al.~2002). These
data, when combined with detailed star formation histories and radial
profiles, will provide significant constraints on the physics of star
formation and baryonic feedback, a major stumbling block in our
understanding of how to model galaxies.  

The modelling of the underlying dark halo needs to go beyond the
obviously inappropriate one-component, mass-follows-light, King models
that have been widely used in the past (e.g.~review of Mateo 1998),
and address issues such as the degeneracy between mass and velocity
anisotropy that are inherent in previous analyses (see Binney \& Mamon
1982 for an elegant discussion of this last point). Recent models that
address these issues for isolated dSph include Wilkinson et al.~(2002)
and Wang et al.~(2005).  However, the dSph are clearly not isolated,
and the time-dependent dynamical effects of their environment must be
taken into account in the modelling, particularly of the outer parts.

\section {The Bulge}

The age distribution of the dominant population of the bulge is
reasonably well-constrained only for Galactocentric z-distances
greater than about 400~pc, Galactic latitude $|b| \simgt 3^\circ$,
since interior to this one faces the complexities of the {\it local\/}
decomposition into bulge and local disk.  Here, analyses of deep HST
colour-magnitude diagrams, after correction of {\it foreground} disk
stars, have shown that the dominant population is old, $\simgt 10$~Gyr
(Feltzing \& Gilmore 2000; Zoccali et al.~2003).  The integrated
population of the inner $10$~degrees of the central regions has been
studied through deep ISO mid-IR data, combined with near IR data from
DENIS, and again the conclusion is that the dominant population is old
(van Loon et al.~2003).  There is also a less significant
intermediate-age component seen in the ISO data, probably that traced
by the OH/IR stars that have been surveyed interior to $|b| = 3^\circ$
(Sevenster 1999).  A young population is also detected by van Loon et
al.~(2003), plus there is ongoing star formation in the plane near the
Galactic Center, most notably traced by populous young clusters (Figer
et al.~1999).  The interpretation of these younger stars in terms of
the stellar populations in the bulge is complicated by the fact that
the scale-height of the thin disk is comparable to that of the central
bulge, so that membership in either component is ambiguous. Continuous
star formation at the present rate, over the last $\sim 10$~Gyr, could
form the entire central stellar cusp, and evidence for this has been
presented (Figer et al.~2004).

 Indeed the relation between the inner triaxial bulge/bar, now failry
well-characterised (see Babusiaux's contribution to this volume, and
Babusiaux \& Gilmore 2005) and the larger-scale bulge is as yet
unclear (see Merrifield 2004 for a recent review).  All that said, the
dominant population in the bulge is probably old.

The metallicity distributions of low-mass stars in various
low-reddening lines-of-sight towards the bulge (with projected
Galactocentric distances of a few 100pc to a few kpc) have been
determined spectroscopically (e.g.~McWilliam \& Rich 1994; Ibata \&
Gilmore 1995; Sadler, Terndrup \& Rich 1996) and photometrically
(e.g.~Zoccali et al.~2003) with the robust result that the peak
metallicity is ${\rm [Fe/H]} \sim -0.3$~dex, with a broad range and a
tail to low abundances (indeed the distribution is well-fit by the
Simple closed-box model, unlike the solar neighborhood data).  The
available elemental abundances, limited to a handful of stars, show
mostly the enhanced ${\rm [\alpha/Fe]}$ signatures of enrichment by
predominantly Type II supernovae (McWilliam \& Rich 1994; McWilliam \&
Rich 2004; Fulbright, Rich \& McWilliam 2005), indicating rapid star
formation.  Indeed the chemical abundances favor very rapid star
formation and (self-)enrichment (Ferreras, Wyse \& Silk 2003).

Deep star counts in the bulge can be used to derive the faint stellar
luminosity function and mass function (modulo mass-to-luminosity
uncertainties; Zoccali et al.~2000).  Comparison with low-metallicity
globular clusters shows that the inferred stellar IMF is invariant
over this range of metallicity.  Further, the faint luminosity
function in the Ursa Minor dSph is also indistinguishable from that of
globular clusters of the same metallicity (Wyse et al.~2002).  The
`Type II plateau' in the elemental abundances of the bulge stars is
equal to that in the stellar halo, implying a fixed massive star
IMF. An invariant IMF is one of the few simplifying assumptions that
we can make with confidence.

The formation of the central bulge then must produce a metal-rich,
predominantly old stellar population, with central regions that
contain a super-massive black hole, and a triaxial bar, with recent
and ongoing star formation.  In the hierarchical clustering scenario,
bulges are built up during mergers, with several mechanisms
contributing.  The dense central regions of massive satellites may
survive and sink to the center; the dynamical friction timescale for a
satellite of mass $M_{sat}$ orbiting in a more massive galaxy of mass
$M_{gal}$ is $t_{dyn\, fric} \sim t_{cross} M_{gal}/M_{sat}$, where
$t_{cross} $ is the crossing time of the more massive galaxy.  With
$t_{cross} \sim 3 \times 10^8$~yr for a large galaxy, only the most
massive satellites could contribute to the central bulge in a Hubble
time.  Gravitational torques during the merger process are also
expected to drive disk gas to the central regions, and some fraction
of stars in the disk will also be heated sufficiently to be
`re-arranged' into a bulge (cf.~Kauffmann 1996).  The predicted age
and metallicity distributions of the stars in the bulge are then
dependent on the merger history; however a uniform old population is
not expected.

An alternative scenario for bulge formation appeals to an instability
in the disk, forming first a bar which then buckles out of the plane
to form a bulge (e.g.~Raha et al.~1991) or is destroyed by the
orbit-scattering effects of the accumulation of mass at its center
(e.g.~Hasan \& Norman 1990).  These secular processes would perhaps
form `pseudo-bulges', identified in observations through, for example,
exponential, rather than $r^{1/4}$, light profiles (e.g.~Carollo et
al.~2001). Again, if this were to be a late-epoch process, with inward
gas flows driven by bars etc., one would expect a significant range of
stellar ages in the pseudo-bulge (see the comprehensive review of
Kormendy \& Kennicutt 2004).

As noted above, the bulge is dominated by old, metal-rich stars.  This
favors neither of the two scenarios above, but rather points to
formation of the bulge by an intense burst of star formation, {\it in
situ}, a long time ago (cf.~Elmegreen 1999; Ferreras, Wyse \& Silk
2003). The inferred star formation rate is $\simgt 10
\cal{M}_\odot$/yr. As noted above, a possible source of the gas is
ejecta from star-forming regions in the halo (see Figure~2 above and
associated discussion).

Or is the bulge really a `pseudo-bulge', formed through secular
instabilities (see discussion in Wyse 1999)? What role is the bar
playing in building the central regions?  What is the connection to
the supermassive black hole at the Galactic Center? Why does the Milky
Way deviate significantly from the correlations of black hole mass with
bulge properties (Tremaine et al.~2002)?

\section {Concluding Remarks}

The mean properties of the dominant stellar populations of the
different components of the Milky Way are reasonably well-defined at a
few locations only.  Even at those locations, we have incomplete
knowledge of the distribution functions.  We do know that the
properties of the different stellar populations overlap, and our
ignorance causes possibly important physics to be lost.

We need to determine detailed age distributions, spatial
distributions, kinematics, elemental abundance distributions etc.~for
large samples of galactic stars, both locally and globally.  We need
further to match the data with robust analyses. Unfortunately,
time-dependent dynamics could be important in several aspects, and
must be addressed -- in the disk, through perturbations from transient
spiral arms and the bar; in the halo and thick disk through tidal
effects and disk `shocking' on substructure; in the bulge through
secular evolution and heating.  We must also push galaxy formation
theories into making more contact with reality, with testable
predictions.

\acknowledgements 
We thank David Valls-Gabaud for his heroic efforts that
resulted in a smoothly running and stimulating meeting (and for moving
the conference closer to Chichen Itza).  RFGW thanks all at the Aspen
Center for Physics for a pleasant and productive environment which
enabled this paper to be written.


\end{document}